\documentclass[%
aps,
superscriptaddress,
twocolumn,
10pt,
showpacs,
showkeys,
floatfix,
pra,
]{revtex4-1}

\usepackage{amsmath,graphicx,cancel,color}
\usepackage{color}
\usepackage{amsthm}
\usepackage{amssymb}
\usepackage{multirow}
\usepackage{nicefrac}
\usepackage[T1]{fontenc}
\usepackage{dcolumn}
\usepackage{bm}
\usepackage{hyperref}
\usepackage[caption=false]{subfig}
\begin{document}
\title{\fontseries{b}\selectfont Students' network integration as a predictor of persistence in introductory physics courses}

\author{Justyna P. Zwolak}
\email[]{j.p.zwolak@gmail.com}
\affiliation{STEM Transformation Institute, Florida International University, Miami, Florida 33199}
\affiliation{Department of Teaching and Learning, Florida International University, Miami, Florida 33199}

\author{Remy Dou}
\affiliation{Department of Teaching and Learning, Florida International University, Miami, Florida 33199}
\affiliation{Department of Physics, University of Maryland, College Park, Maryland, 20742}

\author{Eric A. Williams}
\affiliation{Department of Physics, Florida International University, Miami, Florida 33199}

\author{Eric Brewe}
\affiliation{Department of Physics, Drexel University, Philadelphia, PA 19104}
\affiliation{School of Education, Drexel University, Philadelphia, PA 19104}
\affiliation{STEM Transformation Institute, Florida International University, Miami, Florida 33199}
\affiliation{Department of Teaching and Learning, Florida International University, Miami, Florida 33199}
\affiliation{Department of Physics, Florida International University, Miami, Florida 33199}

\begin{abstract}
Increasing student retention (successfully finishing a particular course) and persistence (continuing through a sequence of courses or the major area of study) is currently a major challenge for universities. While students' academic and social integration into an institution seems to be vital for student retention, research into the effect of interpersonal interactions is rare. We use network analysis as an approach to investigate academic and social experiences of students in the classroom. In particular, centrality measures identify patterns of interaction that contribute to integration into the university. Using these measures, we analyze how position within a social network in a Modeling Instruction (MI) course -- an introductory physics course that strongly emphasizes interactive learning -- predicts their persistence in taking a subsequent physics course. Students with higher centrality at the end of the first semester of MI are more likely to enroll in a second semester of MI. Moreover, we found that chances of successfully predicting individual student's persistence based on centrality measures are fairly high -- up to $75\%$, making the centrality a good predictor of persistence. These findings suggest that increasing student social integration may help in improving persistence in science, technology, engineering, and mathematics fields. 
\end{abstract}
\date{March 10, 2017}
\maketitle

\section{Background and motivation}\label{sec:background}

\begin{figure}[t!]
	\includegraphics[width=0.48\textwidth]{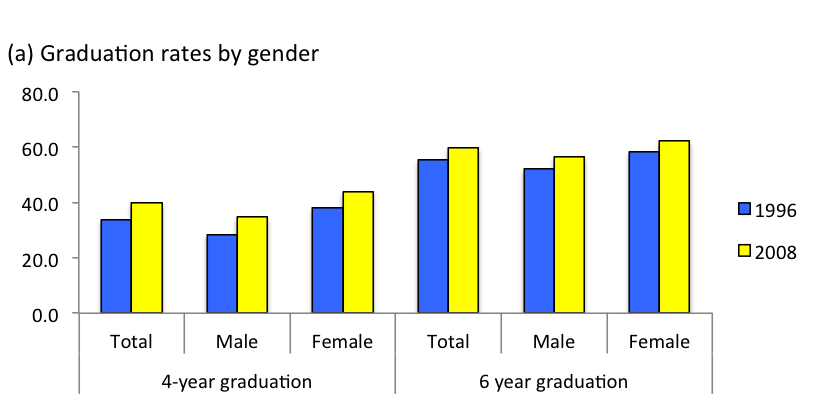}
         \includegraphics[width=0.48\textwidth]{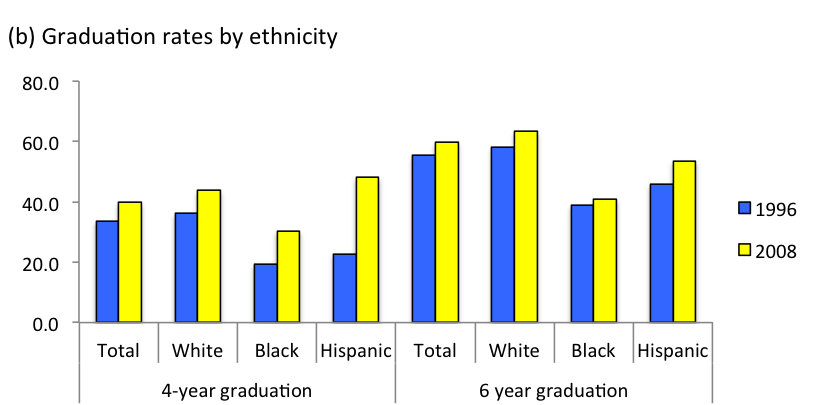}   
    \caption{Graduation rates in 1996 and 2008 (a) by gender and (b) by ethnicity. The 4-year graduation rates increased by about $6\%$, compared to $4\%$ increase of the 6-year graduation rates, for both males and females. By ethnicity, the increase in 4-year graduation rate for Hispanics ($25.3\%$) and Blacks ($10.9\%$) was substantial \cite{grad-rates}.} 
    \label{fig:grad-rates}
\end{figure}

Increasing the retention of students in a particular course and their persistence in continuing through a sequence of courses or their major area of study has always been a big challenge for universities. While postsecondary enrollment has increased tenfold since the 1950s, the institutional graduation rate remained at a constant 50\% level for most of the past half century, increasing only about 10\% over the past two decades (see Fig. ~\ref{fig:grad-rates} for details). In the mid-1990s, the focus of policy makers has moved to the issues of choice, affordability, and persistence. Still, almost half of first-time students who leave their initial institution by the end of the first year never come back to college ~\cite{Swail04-ASR}.

One way to approach this problem is to examine student academic and social integration using the tools of social network analysis (SNA). The basic premise linking networks to persistence is that students' communities and interactions likely influence whether they remain in a particular class, major or in school overall. Network analysis allows us to gain insight into these communities, as well as how they evolve over time. In particular, SNA can be used to identify patterns of interaction that contribute to integration within a classroom and at the university level, and provide quantitative measures to evaluate their importance. It supplies a methodology to assess the structure and quality of interpersonal interactions, both academic and social. For example, centrality measures can reveal students' status within the community, while assortativity coefficients can provide information about their tendency to associate with people of similar background, race, or gender---a manifestation of homophily.

Understanding how embeddedness within the social and academic network of an institution affects students' persistence and retention is crucial for improving their experiences. While students' integration seems to be essential, the implementation of effective practices to prevent losing students is rare. Developing network methodologies for studying retention and persistence among university students---a nascent research area---is a step in this direction~\cite{Brewe12-SNA,Forsman14-CSN}. 

We address students' persistence at Florida International University (FIU)---a large, Hispanic-serving institution. FIU, as is typical for urban universities in major cities, is a commuter school---only 8\% of students live in ``college-owned, college-operated, or college-affiliate'' housing~\cite{FIU-housing}. Among factors that affect student integration into the social and academic environment of the university, one can distinguish between external (e.g., families, neighborhoods, work settings) and internal (e.g., learning groups within a classroom, residence halls) communities~\cite{Tinto75-DHE}. The importance of the classroom as a focal point of interaction is especially pronounced at commuter schools. In particular, introductory courses are likely to strongly influence student persistence. In other words, student success in introductory courses in their first and second years are critical for their continual pursuit of science, technology, engineering, and mathematics (STEM) degrees, including physics~\cite{Tinto97-CAC,Nora03-AHH}.

In our study, we analyze how students' positions within a social network in an introductory mechanics Modeling Instruction (MI-M) course predicts their persistence in the MI sequence [i.e., whether they take the subsequent electricity and magnetism Modeling Instruction (MI-EM) course]. We also look at the associations between persistence and centralities at the beginning of the second semester. Modeling Instruction is a guided-inquiry interactive-engagement method of teaching. It organizes instruction around building, testing, and applying a handful of scientific models that represent the core content of physics. Instead of relying on lectures and textbooks, MI emphasizes students' construction of conceptual and mathematical models in an interactive learning community. It is, therefore, an important case for studying the effects of building communities on promoting persistence~\cite{Brewe10-PLC}.

The paper is organized as follows. We start with a brief overview of persistence studies at the university and classroom levels in Sec.~\ref{sec:lit_rev}. We then present our theoretical framework (Sec.~\ref{sec:framework}), followed by a section on methodology (Sec.~\ref{sec:method}). In particular, the description of the development of the SNA survey is described in Sec.~\ref{subsec:survey}, a short introduction to social network analysis is given in Sec.~\ref{subsec:SNA_terminology}, and data analysis methodology in Sec.~\ref{subsec:data_analysis}. We then move to a discussion of the overall findings. In Sec.~\ref{sec:mi_network} we look at the MI social networks, and, in Sec.~\ref{sec:analysis}, we present the analysis of the relationship between centrality measures and persistence. We conclude with a discussion and potential future research directions in Sec.~\ref{sec:summary}.

\section{Persistence in brief}\label{sec:lit_rev}
\subsection{Persistence and integration}

Although research on undergraduate persistence was conducted as early as the 1930s~\cite{McNeely37-CSM}, it was the publication of Spady's sociological model of student dropout in higher education~\cite{Spady70-DHE}, followed by Tinto's student integration model~\cite{Tinto75-DHE}, that started the current dialogue on undergraduate persistence. In his paper, Spady identified factors that play a part in student social integration and can affect their decision to drop out of school as academic potential, normative congruence, grade performance, intellectual development, and friendship support. He then conducted an empirical study that implicated formal academic performance as the predominant factor for student attrition~\cite{Note1,Spady79-DHE}.

Tinto, on the other hand, suggested that student persistence was linked to both formal and informal academic experiences, as well as social integration. As part of his model, he proposed three principles of effective persistence: (1) institutional commitment to students, (2) educational commitment, and (3) social and intellectual community. In order to be effective, student persistence programs must (1) assure that institutional goals always have a direct or indirect relationship to student success and achievement, (2) commit to the education of all, not just some, of the students, and, finally, (3) help students feel that they are valued and full members of the social and educational communities~\cite{Tinto75-DHE,Tinto87-PER}. Based on the work of Tinto and his followers, increasing students' integration should be one of the prime targets to increase their persistence.

Astin went a step further and reframed the relationship of persistence and involvement into one spectrum, saying, ``if we conceive of involvement as occurring along a continuum'' from least to most involvement, ``the act of dropping out can be viewed as the ultimate form of noninvolvement \dots dropping out anchors the involvement continuum at the lowest end''~\cite{Astin84-SI}. He identified different forms that involvement may take, such as students' place of residence, dedication to academic studies, student-faculty interaction, and participation in extracurricular activities (with special focus on student government, honors programs, and athletics), and found that each of these forms impacts persistence in its own way.

In the 1990s and 2000s, persistence research became more holistic and adopted a multifaceted interdepartmental understanding of how to retain students. This approach invokes a ``cross-departmental institutional responsibility'' for persistence ``via wide-range programming'' that brings together the otherwise-disparate parts of an institution, including admissions officers, instructors, the financial aid office, academic services, and student affairs~\cite{Demetriou11-IMS}. To help students navigate these complex elements of an institution, it became clear that universities must offer accessible support services in a combination of academic, personal, and social contexts in order to support students' persistence~\cite{Tinto04-SRG}. Nora extended Tinto's model to incorporate additional factors, including academic and social integration~\cite{Nora03-AHH}, which was then used to investigate STEM student persistence at a Hispanic-serving institution~\cite{Crisp09-FSP}.

In the context of improving graduation rates of underrepresented minorities, a recent study has found that Hispanic students' ``sense of belonging was positively related to persistence'' in STEM majors, implying that ``greater levels of academic and social integration may be related to higher levels of retention''~\cite{Garcia11-PLP}. However, quantitative research on students' academic and social integration, as well as practical implementations of these findings, are not very common and ``the really difficult work of shaping institutional practice \dots has yet to be tackled''~\cite{Tinto06-RPR}.

The use of social network analysis, which offers ``a new perspective in which integration is expressed as a function of individual social ties'' from students to their peers and instructors, while also incorporating individual background characteristics, makes quantitative research of integration possible~\cite{Thomas00-TTB}. Thomas used this approach to study the link between integration and persistence and found a nuanced relationship between students' social ties and their GPA, goal commitment, and persistence. His work was followed up by several researchers who found a connection between social ties and multiple outcomes, including sense of community and academic performance, which we note explicitly appear as distinct elements within Tinto's integration model~\cite{Dawson08-RSC,Rizzuto09-NWW}. Eckles and Stradley reported that ``factors such as athletic participation, membership in a fraternity or sorority, religion, and ethnicity \dots were not individually significant'' and found that students' persistence from year 1 to year 2 was instead influenced by their friends` persistence~\cite{Eckles12-RAD}. They went on to argue that ``those variables have been significant in the past because they represent strong social connections among like students. They have in effect been working as proxies for social networks \dots membership in such a society puts students in a dense social network that exposes them to more students choosing to stay.'' Hence, it is not only integration in the general sense that matters to a student's persistence, but integration with other persisting students.

\subsection{Classroom as an interaction hub}

Tinto identifies the importance of the classroom, saying that ``what we do not yet know, or at least have not adequately documented, is {\it how} involvement is shaped within the context of differing institutions of higher education by student educational experiences. \dots we have yet to explore the critical linkages between involvement in classrooms, student learning, and persistence.'' He argues that although researchers have not ignored the classroom, their findings remain disconnected from those of the field of student persistence: ``The two fields of inquiry have gone on in parallel without crossing''~\cite{Tinto97-CAC}.

For many individuals, especially new students who have not yet formed connections in the community, the classroom is a place where connecting with others happens. That is even more the case for nonresidential students who have to manage a number of tasks outside of college and the time they spend in class is the only time they spend on campus. Thus, the importance of the classroom experience as a means for improving student persistence must not be understated.

\section{Theoretical framework}\label{sec:framework}

With his model, Tinto approached the issue of attrition from a sociological point of view, emphasizing the importance of integrating new students into the life of the institution, both socially and academically. It was noted by Tinto that ``social and academic life are interwoven and \dots social communities emerge out of academic activities that take place within the more limited academic sphere of the classroom, a sphere of activities that is necessarily also social in character''~\cite{Tinto97-CAC}. However, with the exception of a few studies~\cite{Thomas00-TTB,Dawson08-RSC,Zwolak15-SIP}, persistence at the classroom level remains an open question. In our study, we want to take a fine-grained approach and look precisely at this problem. From a methodological point of view, Tinto suggested that SNA might be a well-suited approach to study the in-class and out-of class interaction: ``\dots we would be well served by \dots network analysis and/or social mapping of student interaction patterns. \dots they will shed important light on how interactions across the academic and social geography of a campus shape the educational opportunity structure of campus life and, in turn, both student learning and persistence''~\cite{Tinto97-CAC}.

\subsection{MI introductory physics classroom}

Given the emphasis on social integration in the persistence literature, physics education researchers have responded by developing active learning courses that take into account modern theories of learning and evidence-based reform~\cite{Beichner00-SU,Etkina01-SLE,Brewe08-MTA,NRC13-ACW}. Modeling Instruction, in particular, shows tremendous promise to increase student outcomes on exams, concept inventories, and attitudes toward physics~\cite{Brewe09-PAS,Brewe10-ETP}. Moreover, while affective outcomes, like self-efficacy, continue to suffer, MI can reduce---and sometimes eliminate---the negative impact on physics self-efficacy when compared to traditional, lecture-based courses~\cite{Dou16-BPM,Sawtelle12-RSR}. One of the most prominent features of MI that sets it apart from lecture-based courses is its adherence to the principle that academic and social interactions between peers and instructors enhance learning.

This principle manifests itself in the classroom through solicited and unsolicited interactions between students and their peers, as well as their instructors, be they a faculty member, a teaching assistant, or a learning assistant. The number of peer-to-peer interactions in MI courses was found to be much higher than in traditional, lecture-based courses. In particular, SNA reveals that networks reported by students in MI sections a few weeks into the semester have a higher density than those reported in traditional sections at the end of the semester (see Table~\ref{tab:net_info} in Sec.~\ref{sec:mi_network} and Table 1 in Ref.~\cite{Brewe10-PLC} for details). In fact, by the end of the semester every student in the MI section had an academic connection with at least one peer, whereas the majority of students in the lecture-based course did not report a single connection~\cite{Brewe10-PLC}. The low level of student interactions in traditional courses indicates that it is difficult to get statistically significant results on the effect of interactions in such courses. This further motivates us to examine student integration in MI courses.

\subsection{Persistence Model}

\begin{figure}[t]
 \includegraphics[width=0.48\textwidth]{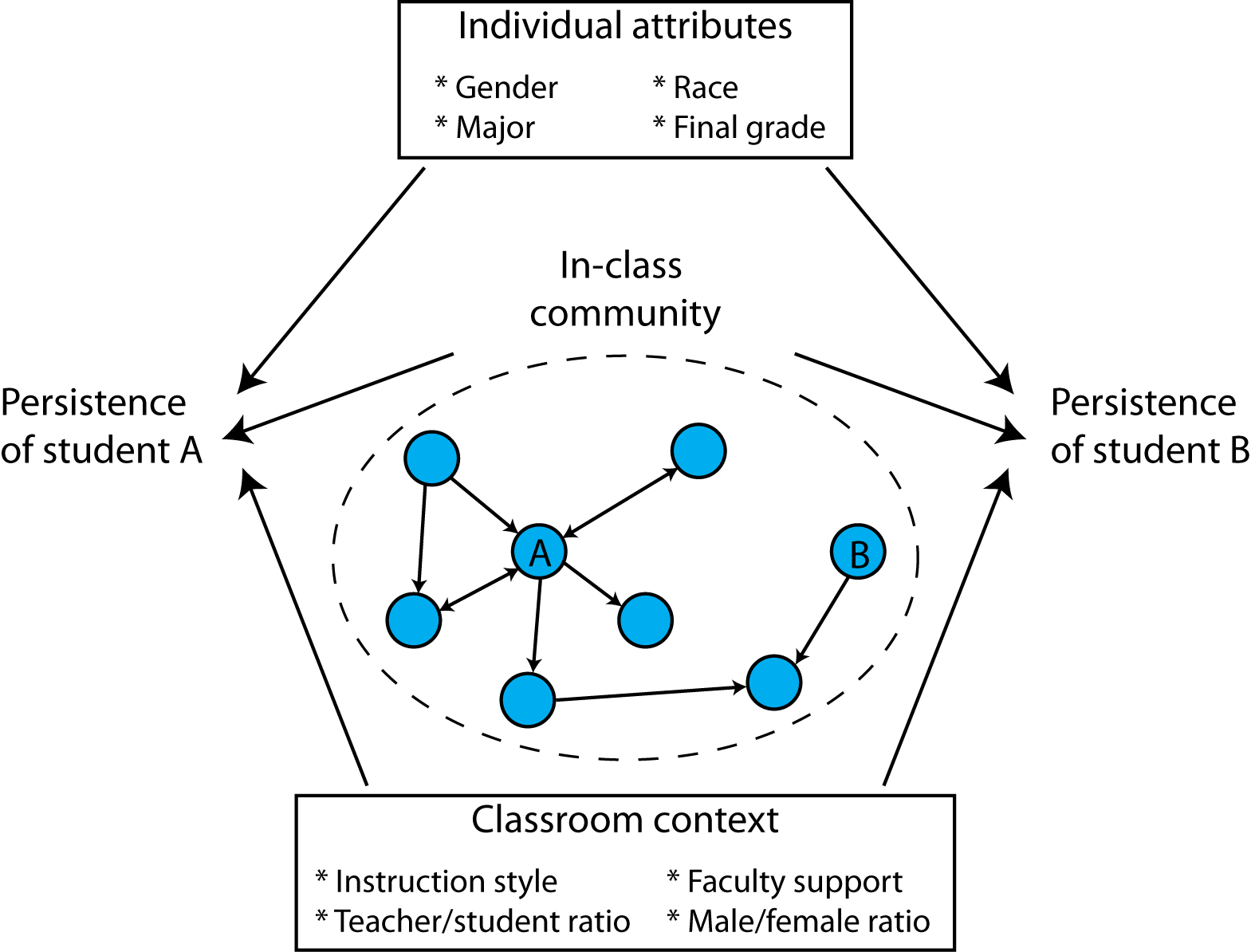}
 \caption{Simplified Model Linking Classrooms, Learning and Persistence. The three categories of factors that may affect students' persistence are individual attributes, classroom context, and in-class community.}
 \label{fig:model}
\end{figure}

Social network research has shown that individual beliefs and behaviors are shaped by social connections and likely do not result from personal attributes alone~\cite{Borgatti03-NPR,Brass95-NPM}. To capture the factors that affect persistence at the classroom level, we propose a simplified version of Tinto's Model Linking Classrooms, Learning and Persistence~\cite{Tinto97-CAC}. Figure~\ref{fig:model} provides a depiction of three categories of factors that we consider. In addition to individual attributes (e.g., gender, race, major) and classroom context (e.g., traditional lecture versus active engagement), we also include the in-class community. The classroom social system may alter students' persistence through better access to resources resulting from interacting with instructional staff and other students (``knowledge access''), peer influence, and/or social and emotional supports. Students who are highly sought by others can be viewed as having high prestige, either academically or socially (or both). Such individuals may hold information or resources that other students deem useful. They may also have personalities that attract others, providing emotional support or becoming academic or social ``role models.'' Individuals with positive attitudes towards sciences---individuals who are excited and passionate about the subject---may get other students more interested and engaged in the course. On the other hand, students who are not necessarily resourceful but who are perceived by others as ``strong personalities'' can draw them away from sciences by expressing lack of interest in the subject matter. Shared beliefs are developed through interaction and exposure to the beliefs of others~\cite{Ibarra93-PSS}. Peer interactions are necessarily affected by the levels and types of social capital that students possess, especially in a collaborative setting~\cite{Claridge04-SCM}. In a classroom with emphasis on peer interaction as the process by which learning occurs, such as Modeling Instruction, a social network perspective offers a valuable lens for investigating the association between peer interactions and students' persistence~\cite{Brewe08-MTA}.

\section{Methodology}\label{sec:method}
\subsection{The SNA survey}\label{subsec:survey}

\begin{figure}[t]
 \includegraphics[width=0.48\textwidth]{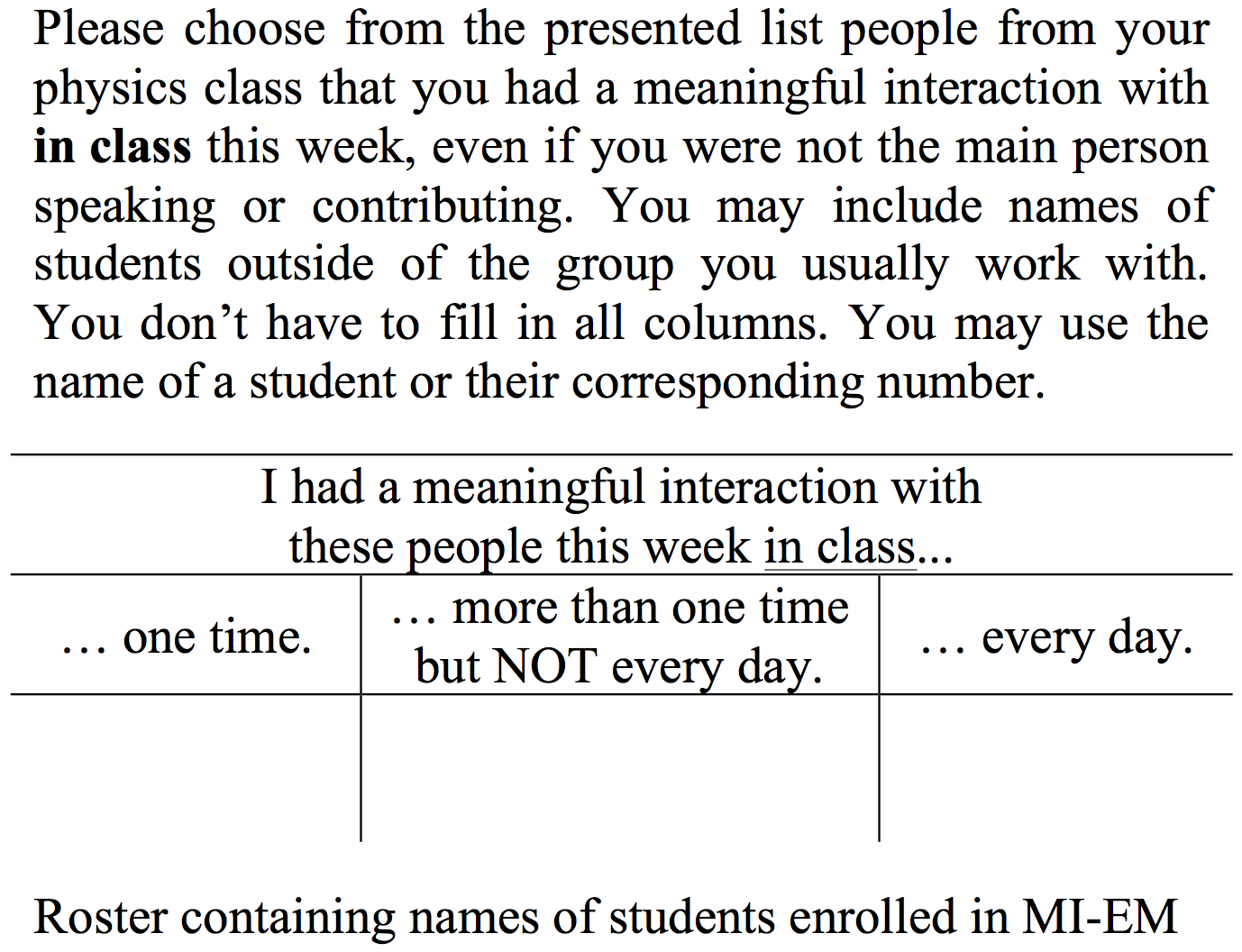}
 \caption{An excerpt from the SNA survey that was given starting in the Spring 2015 semester.}
\label{fig:SNAsp15}
\end{figure}

To collect social network data we have developed a pencil and paper survey that was administered in the introductory mechanics MI course (Fall 2014, Fall 2015) and in the electricity and magnetism MI course (Spring 2015, Spring 2016). The survey was given every 3--4 weeks throughout the semester. Students were asked the following question:
\begin{quote}
Name the individual(s) (first and last name) you had a meaningful classroom interaction* with today, even if you were not the main person speaking or contributing. \textit{(You may include names of students outside of the group you usually work with)} \\
\small
*A classroom interaction includes but is not limited to people you worked with to solve physics problems and people that you watched or listened to while solving physics problems.
\normalsize
\end{quote}
In the pilot data collection (Fall 2014), we used a simplified version of the survey that consisted of the question followed by a blank space where students were supposed to write names of their peers. Starting in Spring 2015, we switched to a version of the survey that included a roster with names of all students enrolled in the course in a randomized order and of the instructional staff. Moreover, since not all interactions are equally meaningful, we decided to introduce a weighted version of the question about interactions, as shown in Fig.~\ref{fig:SNAsp15}. It has been noted that ``knowing that someone else has valuable expertise is important, but their knowledge is really helpful only if they are accessible''~\cite{Borgatti03-RVN}. Thus, it is important to go beyond just the number of ties one has in the network and look also at their ``quality.''

Since all interactions listed are considered ``meaningful'' (by the phrasing of the SNA question), in our context the quality of ties is represented by the frequency of their occurrence---if student $A$ reported daily meaningful in-class interactions with student $B$, we considered it to be more important (of ``higher quality'') than a one-time interaction. To express the importance of ties in the language of networks, we assigned a numeric value to the frequency of interaction by weighting them from $0$ to $3$ for any given pair of students, with weight $0$ being assigned to an unreported interaction. For example, if student $A$ reported having an in-class interaction with student $B$ every day, $w_{AB}=3$, and if student $B$ did not mention student $A$ on a survey, $w_{BA}=0$.

Because of the intrinsically interactive nature of the relational data, it is of particular importance to define the network boundaries. Not including some relevant nodes or ties may affect the properties of both the entire network and those of each individual node. In our case, the exogenously determined boundaries are defined by the enrollment in the Modeling Instruction course.

\subsection{Quantifying social integration}\label{subsec:SNA_terminology} 

\begin{figure}[t]
 \centering
 \includegraphics[width=0.45\textwidth]{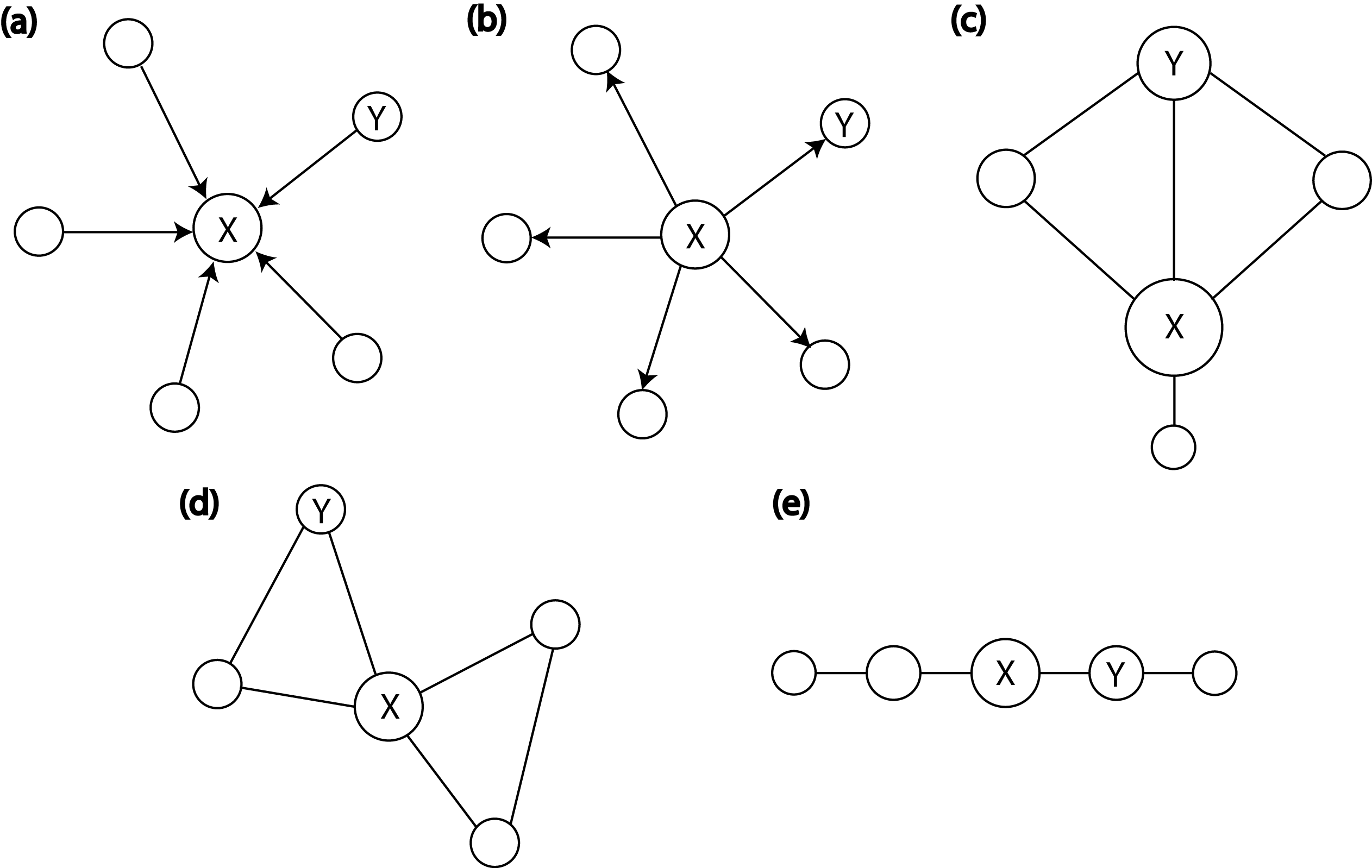}
 \caption{In each of the networks above, $X$ has higher centrality than $Y$ according to (a) indegree, (b) outdegree, (c) eigenvector, (d) betweenness and (e) closeness}.
 \label{fig:centralities}
\end{figure}

SNA uses the notion of nodes (in our case students enrolled in MI-M) and edges (the interactions identified by students on the survey) to represent the network. From a graph theoretic perspective, the relative importance of nodes within a graph is determined using centrality measures. Evaluating the centrality of nodes in the network helps us to understand the network and its participants (see Fig.~\ref{fig:centralities}) and to answer the question: ``Who are the most important nodes in a network?''~\cite{Note2}.

\subsubsection{Centrality measures}

There are various measures of centrality that quantify the importance of nodes and edges. In this paper we will focus on three groups of measures: node-level centralities (such as directed degrees), the whole network measures (such as betweenness and closeness), and measures that bridge a gap between these two extremes (such as eigenvector). A brief description of these measures for an unweighted network, followed by a short introduction to centrality normalization, is presented below (see, e.g., Refs.~\cite{Wasserman1994,Scott2011,Prell2011}).

{\it Degree} (also called {\it total degree}) centrality is the number of edges directly connected to a given node. It can be thought of as a measure of connectivity. In the case of a directed network, i.e., a network that takes into account the origin of an edge, one can define two directional measures of degree centrality: {\it indegree}---the number of ties directed to the node (can be interpreted as popularity)---and {\it outdegree}---the number of ties that the node directs to others (can be interpreted as sociability or influence):
\begin{equation*}
C_{D}^{\leftarrow}(i)=\sum_{j=1}^n x_{ji} \hspace{0.2in},\hspace{0.2in} C_{D}^{\rightarrow}(i)=\sum_{j=1}^n x_{ij}.\vspace{-0.10cm}
\end{equation*}
where $x_{ji}$ is $1$ when there is a tie from node $j$ to node $i$ and $0$ otherwise (sum of all $i$'s incoming ties) and $x_{ij}$ is $1$ when there is a tie from node $i$ to node $j$ and $0$ otherwise (sum of all $i$'s outgoing ties).

{\it Eigenvector} centrality goes beyond the node in question and looks also at the centrality of the nodes connected to it. It is defined as the sum of a node's connections to other nodes weighted by their degrees and it measures the influence of a node in a network. It is given by an eigenvector of an adjacency matrix $\boldsymbol{A}$ corresponding to the greatest eigenvalue $\lambda_{max}$. That is,\vspace{-0.05cm}
\begin{equation*}
\boldsymbol{A}^T\vec{\boldsymbol{C}}_E=\lambda_{max}\vec{\boldsymbol{C}}_E.\vspace{-0.05cm}
\end{equation*}
Here, $\boldsymbol{A}$ is the adjacency matrix representation of a graph, such that $a_{ij} = 1$ if a node $i$ is connected to a node $j$ by an edge and $0$ if it is not. Then, $\vec{\boldsymbol{C}}_E$ is a vector containing the centralities of all nodes in the network.

Directed degrees as well as eigenvector centrality are relatively intuitive and easy to calculate. They are local measures, and the network outside of the vicinity of a node---i.e., outside the ``ego network''---has no direct influence on them~\cite{Note3}. To assess the effect of the entire network on a given node, one needs a global, system-dependent measure, that will also account for the impact of the nondirectly connected nodes. The following two centralities serve this purpose.

{\it Betweenness} quantifies the number of times a node acts as a bridge along the shortest path linking two other nodes (the {\it geodesic}). It captures the importance of a position within a whole network and can be interpreted as a measure of how much control over the flow of information, and as a consequence how much influence over the entire network, a node has. When there is only one path connecting two nodes or if a node falls on all geodesics connecting two nodes, then it has complete control over the communication between the two other nodes. If, however, a node lies on some but not all geodesics connecting two other nodes, its potential for control is more limited and is proportional to the number of geodesics that a node lies on. Betweenness is given by
\vspace{-0.05cm} 
\begin{equation*}
C_B(k)=\sum_{i\neq j \neq k}^n \frac{d_{ij}(k)}{d_{ij}},\vspace{-0.05cm}
\end{equation*}
where $d_{ij}(k)$ is the number of shortest paths linking node $i$ to node $j$ that pass through node $k$, $d_{ij}$ the number of shortest paths linking node $i$ to node $j$.

Finally, \textit{closeness} is a measure of how near an individual is to all other nodes in a network. It emphasizes a node's independence -- a node that is close to many other nodes can easily reach others without having to rely much on intermediaries, thus gaining easy access to information or resources in the network. It is defined as the inverse of the sum of distances from all other nodes: \vspace{-0.05cm}
\begin{equation*}
C_C(i)=\bigg[\sum_{j=1}^n d_{ij}\bigg]^{-1},\vspace{-0.05cm}
\end{equation*}
where $d_{ij}$ is the geodesic connecting node $i$ to node $j$. When there is no path between two nodes the total number of nodes is used in the formula instead of the path length.

\subsubsection{Different size networks}

To compare measures between different graphs, one needs a measure from which the effect of network size has been removed. To do so, we have to transform all the values to fall within the $[0, 1]$ by dividing them by the highest possible value for each measure~\cite{Wasserman1994}. Since in a network of size $n$ a given node can be in direct contact with at most $n?1$ other nodes, the normalization factor for directed degrees is $1/(n-1)$. That is,\vspace{-0.10cm}
\begin{equation*} 
\Big[C_{D}^{\nicefrac{\leftarrow}{\rightarrow}}\Big]^{nor}(i)=\frac{C_{D}^{\nicefrac{\leftarrow}{\rightarrow}}(i)}{n-1} .\vspace{-0.10cm}
\end{equation*} 
Similarly, since closeness is based on a distance of a given node from $n-1$ other nodes, it is given as\vspace{-0.10cm}
\begin{equation*}
C_C^{nor}(i)=(n-1)\,C_C(i).\vspace{-0.10cm}
\end{equation*} 
For betweenness the normalization factor is given by the maximum value that $C_B(k)$ can take, which is is $\frac{(n-1)(n-2)}{2}$ (true for a star graph, see Fig. \ref{fig:centralities}a and \ref{fig:centralities}b). Therefore\vspace{-0.15cm}
\begin{equation*}
C_B^{nor}(k)=\frac{2\,C_B(k)}{(n-1)(n-2)}.\vspace{-0.10cm}
\end{equation*}

\subsection{Demographics}\label{subsec:demographics}

Data collection for this study occurred at a large public research university over two semesters of a MI-M course (Fall 2014, Fall 2015) and two semesters of a MI-EM course (Spring 2015, Spring 2016), spanning three sections of MI-M and three sections of MI-EM (see Table~\ref{tab:stud-enrol} for details). The sections were taught by two instructors (denoted in Table~\ref{tab:stud-enrol} as $A$ and $B$, both physics education researchers) accompanied by teaching assistants (graduate students in physics education) and learning assistants (high-achieving undergraduate students who have taken the course previously).
\begin{table}[b]
\renewcommand{\arraystretch}{1.1}
\renewcommand{\tabcolsep}{3pt}
\caption{Student enrollment and teaching staff for the MI courses in numbers. There was one section of MI-M in Fall 2014, one section of MI-EM in Spring 2015, two sections of MI-M in Fall 2015 and two sections of MI-EM in Spring 2016. Sections $A$ and $B$ were taught by the same instructors. The LAs and TAs varied from semester to semester.}
\centering
\begin{tabular}{lcrrcrr}
\hline \hline
 & Fall 2014 & \multicolumn{2}{c}{Fall 2015} & Spring 2015 & \multicolumn{2}{c}{Spring 2016} \\ \hline
Section & A & A & B & A & A & B \\
Instructors & $2^*$ & $1$ & $1$ & $1$ & $1$ & $1$\\
Students & $73$ & $73$ & $74$ & $73$ & $76$ & $68$\\ 
TAs & $2$ & $1$ & $2$ & $2$ & $1$ & $2$\\ 
LAs & $3$ &$3$ & $3$ & $2$  & $2$ & $3$\\ 
\hline \hline 
\multicolumn{7}{c}{\multirow{2}{8.5cm}{\footnotesize $^*$There was one instructor teaching the course in Fall 2014 and one faculty member who visited the class throughout the semester.}}
\end{tabular}
\label{tab:stud-enrol}
\end{table}

Because of its structure, there is a well-defined plan to follow for the Modeling Instruction courses. Instructors use well-developed resources and have weekly preparation meetings with the entire instructional staff to assure the consistency of teaching the core concepts~\cite{Note4}.

In each semester we collected SNA data five times throughout the duration of the course. The response rates on all surveys but one were $78\%$ or more (see Table~\ref{tab:response-rate}). Therefore, we disregarded the survey with an unusually low return ($43\%$) from the analysis. Moreover, to prevent the low response rate on the last survey, we rescheduled the data collection starting in Spring 2015 onward~\cite{Note5}. As a result, the fifth data collection in Fall 2015 took place around the same time during the semester as the fourth collection in Fall 2014 (fourth week of November). Thus, in our analysis we are using SNA4 (the last valid survey) from the Fall 2014 semester and SNA5 for Fall 2015.

Because of a limited capacity of the MI classroom (up to 80 students per section), students enroll in the course in the order they sign up for it. If the number of students exceeds the number of available spots, a lottery system is utilized in order to distribute permits fairly. The total number of students enrolled in the MI-M was 220, 96 of which were female and 124 were male, and 218 for MI-EM (82 females, 136 males). Both sections of MI were taken by 148 students (64 females, 84 males) while a second semester of physics in a more traditional arrangement was taken by 13 students from MI-M (6 females, 7 males). The ethnicity distribution is provided in Table~\ref{tab:ethn}.
\begin{table}[t]
\renewcommand{\arraystretch}{1.1}
\renewcommand{\tabcolsep}{3.5pt}
\caption{Response rates to the Fall 2014 and Fall 2015 SNA surveys. The unusually low return on survey SNA5 from Fall 2014 was likely due to the optional nature of the review session class when the data collection took place. These data were therefore disregarded from the analysis.}
\centering
\begin{tabular}{lccc}
 \hline \hline
Collection &  Fall 2014 & Fall 2015 (Sec.A) &  Fall 2015 (Sec.B)   \\ \hline  
SNA\#1 &  95         & 97 & 96 \\
SNA\#2 &  92         & 86 & 84 \\
SNA\#3 &  78         & 79& 89 \\
SNA\#4 &  80         & 83 & 80 \\
SNA\#5 &  43$^*$ & 79 & 86 \\ 
\hline \hline
\label{tab:response-rate}
\end{tabular}
\end{table}

\subsection{Data analysis methodology}\label{subsec:data_analysis}

To investigate relationships between students' centralities, gender, ethnicity, major of study, final grade, and their persistence in MI, logistic regression modeling was used. Our outcome variable was persistence through the MI introductory course sequence as measured by student's enrollment (i.e., 1) or lack thereof (i.e., 0) in MI-EM during the subsequent semester. To avoid confounding factors, we performed multiple logistic regression. All significant variables for the simple linear regression analysis were incorporated into the expanded model. The comparison of the fit of simple and multiple linear regression models was performed using the likelihood ratio test, with the null hypothesis stating that the simple model is a better predictor of persistence. The variance inflation factor (VIF) was examined for each of the predictor variables to test the multicollinearity within the model. Variables with a VIF greater than 2.5 were excluded from the final model~\cite{Note6}. Finally, the mutual information approach was used to find the most significant split into the predicting and nonpredicting categories for each of the centrality measures and the chi-square test was used to verify significance of this split~\cite{Note7}. For the statistical analysis we used the R statistical programming language~\cite{R}, and for network analysis we used the igraph package~\cite{igraph}. To adjust the false discovery rate the Benjamini-Hochberg procedure was implemented~\cite{Benjamini95-CFD}. We considered results with $p < 0.05$ as significant.

Starting in Spring 2015, we provided students with a roster of all students enrolled in the class as they responded to the survey. This led to nearly doubling of the number of reported ties per person. In order to aggregate networks from two different semesters, we need them to be similar in terms of various characteristics (see Sec.~\ref{sec:mi_network} for details). In Fall 2014 we asked students about interactions ``today,'' while on the Fall 2015 survey interactions with weight 3 were defined as ``everyday in-class interactions.'' Thus, we decided to include in our analysis for the latter only ties with weight 3 so as to compare ties of approximately equal meaning. It is important to note that this step still yielded about $22\%$ more interactions than was reported in Fall 2014, which is reflected in network characteristics, e.g., slightly higher density, lower average path length, and diameter (see Table~\ref{tab:net_info} for details).

\begin{table}[t]\renewcommand{\arraystretch}{1.1}
\renewcommand{\tabcolsep}{6pt}
\caption{Students' ethnicity distribution. For each group, the first column gives the overall number of students enrolled and the second column gives the average percentage of students between the three sections.}
\centering
\begin{tabular}{p{3.3cm}p{2.0cm}p{2.0cm}}
 \hline \hline
 &  $N$ & Mean [\%] \\ \hline  
Asian &  	$14$    & \hspace{1.5mm}$6.4$   \\
Black &  	$22$    & $10.0$  \\
Hispanic &  $156$ 	& $71.0$ \\
White &  	$22$ 	& $10.0$ \\
Other/NA &  $6$ 	& \hspace{1.5mm}$2.7$  \\
\hline \hline
\label{tab:ethn}
\end{tabular}
\end{table}

\begin{table*}
\renewcommand{\arraystretch}{1.1}
\renewcommand{\tabcolsep}{6pt}
\caption{Comparison of network characteristics for the Fall 2014 (SNA4), Fall 2015 (SNA5, two sections), Spring 2015 (SNA1) and Spring 2016 (SNA1, two sections): network size ($n$), density ($\Delta$), outdegree centralization ($C_{d}^{\rightarrow}$), closeness centralization ($C_{c}$), average path length ($L$), diameter ($D$), transitivity ($Tr$), and reciprocity ($\rho^{\leftrightarrow}$).}
\centering
\begin{tabular}{p{1.2cm}p{0.8cm}p{1.0cm}p{1.2cm}p{1.2cm}p{0.8cm}p{0.8cm}p{1.0cm}p{1.0cm}} \hline \hline
  & $n$ & \hspace{1.0mm}$\Delta$  & \hspace{0.5mm}$C_{d}^{\rightarrow}$ & \hspace{0.5mm}$C_c$ & \hspace{0.5mm}$L$ & $D$ & \hspace{0.5mm}$Tr$ & \hspace{0.5mm}$\rho^{\leftrightarrow}$\\ \hline
F14           & $80$ & $0.05$ & $0.093$ & $0.021$ & $4.6$ & $12$ & $0.42$ & $0.58$ \\
F15$_{Sec.A}$ & $78$ & $0.07$ & $0.088$ & $0.029$ & $3.2$ &  \hspace{1.5mm}$9$ & $0.31$ & $0.42$ \\
F15$_{Sec.B}$ & $80$ & $0.07$ & $0.087$ & $0.088$ & $3.7$ &  \hspace{1.5mm}$9$ & $0.35$ & $0.46$ \\ \hline
S15           & $79$ & $0.06$ & $0.069$ & $0.038$ & $4.2$ & $12$ & $0.44$ & $0.59$ \\
S16$_{Sec.A}$ & $80$ & $0.07$ & $0.107$ & $0.054$ & $3.6$ & $10$ & $0.37$ & $0.48$ \\
S16$_{Sec.B}$ & $74$ & $0.06$ & $0.117$ & $0.065$ & $2.7$ &  \hspace{1.5mm}$9$ & $0.41$ & $0.48$ \\
\hline \hline
\end{tabular}
\label{tab:net_info}
\end{table*}

\subsection{Handling missing data}\label{subsec:missing_data}

The SNA data collection took place in the classroom, at the end of a class. As a result, none of the surveys has a $100\%$ response rate since on any given day some of the students were not present and others had to leave the classroom before the end of class. Response rates for all the surveys are presented in Table~\ref{tab:response-rate}. 

One way to avoid missing values when calculating centralities is to include in a network only those students who either took a given survey or were listed by their peers and then to impute all missing centralities. Another way is to treat all students enrolled in the course as members of a network on each collection and calculate centralities for all of them based on the available data. With this approach students' whose names did not appear in a given collection will naturally have assigned centrality values corresponding to ``isolates'' (disconnected members of a network). It is important to note that changing a centrality for one node can have an effect on centralities of many or all other members of the network, depending on the measure in question. Imputation fills in the missing values without changing values already calculated. While it can be a good approach for handling interdependent missing data when the fraction of unavailable data is small, in our case the missing data account for about 20\% of the data and there is a risk that imputation would significantly change the properties of the network. At the same time, centrality scores are fairly robust to random missingness. For example, for small networks (40--75 nodes) the level of missing data that does not affect the overall structure is up to 35\% for directed degrees and about 20\% for closeness and betweenness~\cite{Smith13-SES}. The missingness in our network data falls within these thresholds and therefore no imputation was used.

\section{The Modeling classroom network}\label{sec:mi_network}

In our analysis we considered six classroom networks: SNA4 from Fall 2014, SNA5 from Fall 2015 (sections A and B) and SNA1 from Spring 2015 and Spring 2016 (sections A and B). Centrality measures were calculated separately for each of these sections and then the resulting indices, together with student demographics, as well as information about grades and persistence, were aggregated into two data sets---one for Fall and one for Spring. To justify the aggregation of data from three different sections, we took a closer look at the properties of each network (see Table~\ref{tab:net_info} for details).

\subsection{Network density}

\begin{figure}[b]
 \centering
 \includegraphics[width=0.48\textwidth]{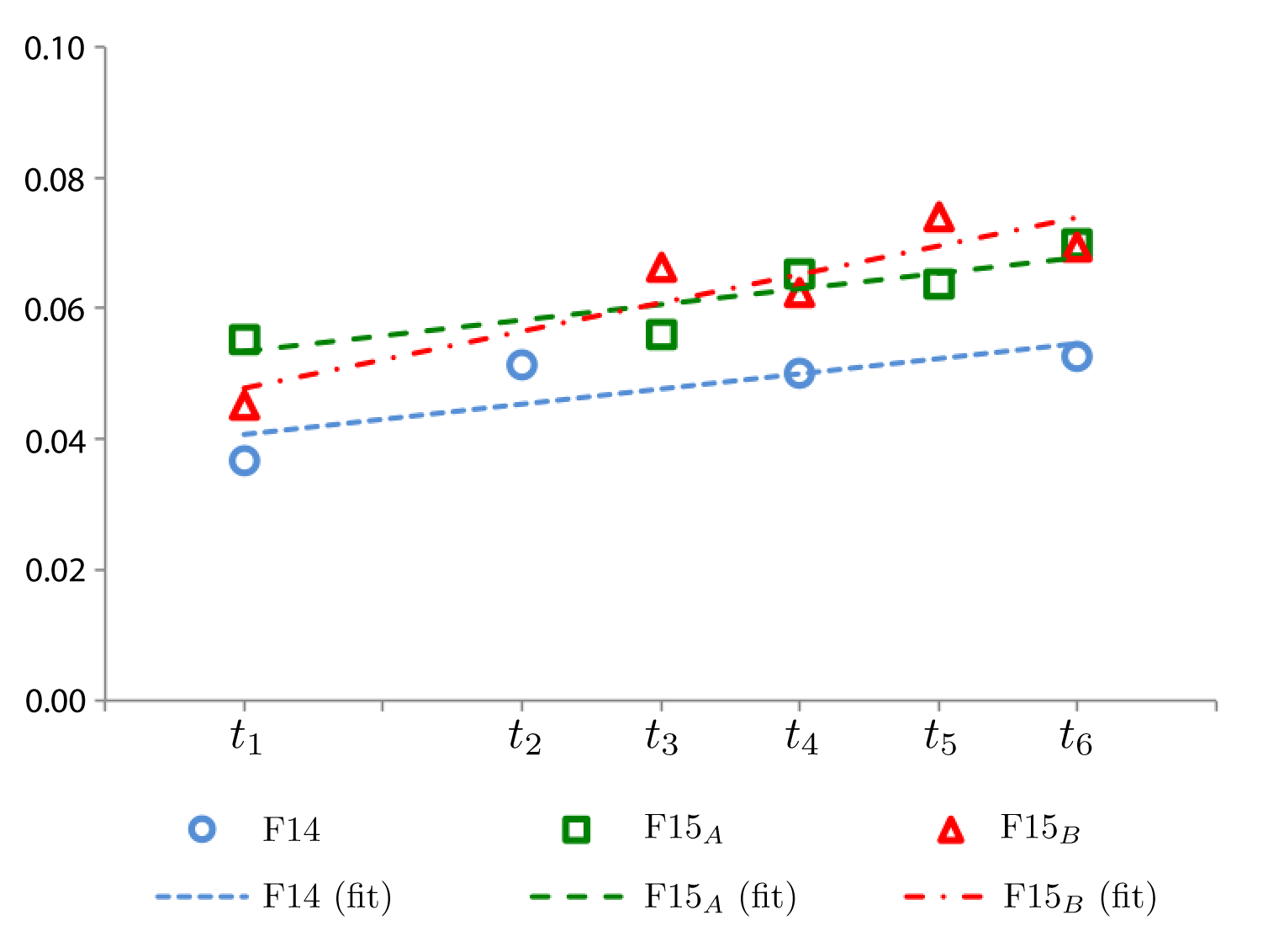}
 \caption{The changes in network density for each section as a function of time. The $x$-axis is re-scaled to account for the adjustment on the data collection schedule. The solid lines represent a line of best fit for each section ($R^2_{F14}=0.65$, $R^2_{F15_A}=0.76$, $R^2_{F15_B}=0.83$).} 
 \label{fig:net_dens}
\end{figure}

For graph densities, we expected that, as the semester progresses, students would get better acquainted with one another and, as a result, the number of reported ties, and therefore also the network density, would systematically increase, leading to faster information transmission and greater access to resources and peer support (see, e.g., Ref.~\cite{Prell2011}). Based on our theoretical framework, students embedded in dense classroom networks should be more likely to persist due to an overall greater exposure to the advantages of peer learning as well as broader range and accessibility of available knowledge resources and social and academic support. For a directed network, density is defined as \vspace{-0.15cm}
\begin{equation*}
\Delta=\frac{\ell}{n(n-1)},\vspace{-0.05cm}
\end{equation*}
where $\ell$ is the number of all ties in a network and $n(n-1)$ is the number of all possible directed ties between $n$ nodes, and it takes value between 0 and 1~\cite{Wasserman1994}. The changes of the network density throughout the semester for each section are presented in Fig.~\ref{fig:net_dens}. One can see that not only the overall trend for all graphs is positive, indicating that the density within each section indeed increased, but also the densities on last collections are comparable. The slightly lower density for Fall 2014 is most likely due to a different format of the survey. For the Spring data the densities were more uniform with $M_\Delta=6.37\cdot10^{-2}$ ($SD_\Delta=0.01\cdot10^{-2}$).

\subsection{Centralization}

While density describes the general level of cohesion in a graph, centralization describes the extent to which this cohesion is organized around particular focal points. In other words, centralization measures how much variation there is in the centrality scores among nodes. It is calculated by looking at the differences between the centralities of the most central node and those of all other nodes and then finding the ratio of the actual sum of differences to the maximum possible sum of differences. High centralization values (close to 1) indicate that there are dominating nodes in the network while low values (close to 0) indicate relatively equal distribution on centrality measures among nodes~\cite{Wasserman1994}. A summary of outdegree ($C_{d}^{\rightarrow}$) and closeness ($C_c$) centralizations for all surveys is presented in Table~\ref{tab:net_info}. These summaries indicate fairly uniform distributions for both measures. The corresponding values for the remaining centralities are similarly low and comparable.

\subsection{Diameter and average path length}

Another way to compare network cohesion is through the network diameter and average path length. Diameter is the length of the longest path between two nodes. It provides information about the span of a network. The average path length, on the other hand, is the shortest path between two nodes, averaged over all pairs of nodes. It is an indicator of how close together nodes are to one another~\cite{Wasserman1994}. Both diameter and average path length are considered to be very robust measures of network topology~\cite{Albert03-SMN}. The average path length will be bounded above by the diameter and is usually much shorter than the diameter. This holds in our case. When normalized by the network size, both measures for each network are comparable, with $M_{\nicefrac{L}{N}}=0.36$ ($SD_{\nicefrac{L}{N}}=0.04$) and $M_{\nicefrac{D}{N}}=0.13$ ($SD_{\nicefrac{D}{N}}=0.02$)

\subsection{Reciprocity and transitivity}

Reciprocity is a tendency of pairs of nodes to form mutual connections between each other. Transitivity, on the other hand, refers to the extent to which the relation between two nodes is transitive, i.e., two connected nodes have a common neighbor (``a friend of my friend is also my friend'')~\cite{Wasserman1994}. Both measures take values from the interval $[0,1]$. As shown in Table~\ref{tab:net_info}, in the case of the MI networks both reciprocity and transitivity are relatively high at the beginning and at the end of the semester ($M_{Tr}=0.38$, $SD_{Tr}=0.05$ and $M_{\rho^\leftrightarrow}=0.49$, $SD_{\rho^\leftrightarrow}=0.05$). That is likely due to the nature of the MI structure (i.e., working in groups of three, sitting at tables of six).

\section{Analysis and results}\label{sec:analysis}

The leading research questions for our study were as follows:
\begin{enumerate}
\item How does students' position within a social network in an MI-M course, which strongly emphasizes interactive learning, impact their persistence in taking a subsequent physics course?
\item Does participation in the Fall MI course tell us something about students' position within the classroom network at the beginning of the Spring MI course?
\end{enumerate}

\subsection{Centrality as a predictor of persistence}\label{subsec:centrality_as_predictor}

In our analysis, we are interested in the effect of student embeddedness within the classroom network on their persistence in the MI sequence. In our preliminary study, we looked at the networks without the instructional staff as we were interested mainly in the peer-to-peer interactions~\cite{Zwolak15-SIP}. However, since the instructors can also be a source of both academic and social support, in our model we decided to take into account the effect on instructors on the network. Thus, we consider two cases: (1) all network interactions as reported and (2) interactions between students excluding the instructional staff. Using the Wilcoxon rank-sum test we found no evidence for statistically significant differences between the two population medians (i.e., with and without instructors) for one out of five centralities we considered---the outdegree. However, the predictive power for all of them remained unchanged, regardless of whether the instructional staff was excluded or not.

For the simple logistic regression models, i.e., persistence $\sim$ centrality, the independent variable is continuous and the dependent variable is binary (true or false). As shown in Table~\ref{tab:centr-all}, we found statistically significant positive correlations for persistence in MI with directed degrees, as well as closeness.

\begin{table}[t]
\renewcommand{\arraystretch}{1.1}
\renewcommand{\tabcolsep}{6pt}
\caption{Estimates for the simple logistic regression for persistence as predicted by various centrality measures (persistence $\sim$ centrality). We consider networks with and without instructional staff (full and student network, respectively). Significant adjusted $p$ values are marked with an asterisk.}
\centering
\begin{tabular}{p{2.5cm}p{2.4cm}p{2.4cm}} \hline \hline
Centrality & Full network & Student network \\ \hline
Indegree & \hspace{0.5mm}$23.84^{**}$ & \hspace{0.5mm}$21.93^{**}$ \\ 
Outdegree  & \hspace{0.5mm}$14.99^{**}$ & \hspace{0.5mm}$15.91^{**}$ \\
Eigenvector & \hspace{2.0mm}$1.22$ & \hspace{2.5mm}$0.43$  \\  
Betweenness  & \hspace{2.0mm}$6.35$ & $-2.59$ \\ 
Closeness & \hspace{2.0mm}$8.64^{***}$ & \hspace{2.5mm}$6.96^{**}$  \\ 
 \hline \hline
\multicolumn{3}{l}{\footnotesize ***$p<0.001$, **$p<0.01$, *$p<0.05$}
\end{tabular}
\label{tab:centr-all}
\end{table}

To determine whether our simple models can be improved, we considered multiple logistic regression models for all statistically significant centrality measures, with a student's gender, ethnicity, academic plan (declared major), and a final grade considered as additional predictors, i.e., persistence $\sim$ centrality + gender + ethnicity + major + grade. To account for other factors as possible predictors, we used a backward elimination regression approach. In particular, to fit the best models to our data we used the \texttt{step} function in R, relying on Akaike's Information Criterion. Starting with a full model, including all candidate variables, and using a comparison criterion to test the removal of variables (i.e., removing the variable that leads to the best improvement in the model and repeating until no further deletion can be performed), we ended up with two-predictors models. We found that only grade made a statistically significant difference in the model fits. Table~\ref{tab:nested-models} summarizes the results of the logistic regression. However, when we compared the fit of the full models (i.e., models with both predicting variables) to the fit of the models with a grade as a sole predictor, we found that grade alone gave a significantly better fit only for indegree [$\chi^2(1)=0.31$, adjusted $p=0.62$]. Full models remained better fits for outdegree [$\chi^2(1)=4.56$, adjusted $p=0.048$] and closeness [$\chi^2(1)=8.25$, adjusted $p=0.007$]. The variance inflation factor indicates the lack of collinearity between grade and both outdegree ($VIF=1.00$) and closeness ($VIF=1.01$).

Finally, to optimize the correlation and to determine the predictability threshold for centralities, we used the mutual information. Table~\ref{tab:mutual-inf} shows the threshold values for each centrality measure and its significance level, as well as probabilities of successfully inferring the persistence based on a given centrality.

\begin{table}[t]
\renewcommand{\arraystretch}{1.1}
\renewcommand{\tabcolsep}{6pt}
\caption{Summary of the likelihood ratio test performed for the multiple logistic regression models with a student's final grade considered as additional predictor (persistence $\sim$ centrality + grade) when compared to the simple models (persistence $\sim$ centrality). Significant adjusted $p$ values are marked with an asterisk.}
\centering
\begin{tabular}{p{3.3cm}p{2cm}p{2cm}}
\hline \hline
Centrality & d.o.f. &  $\hspace{2mm}\chi^2$ \\ \hline
Indegree & \hspace{0.5mm}$1$ & \hspace{0.5mm}$52.9^{***}$ \\ 
Outdegree  & \hspace{0.5mm}$1$ & \hspace{0.5mm}$55.3^{***}$ \\
Closeness & \hspace{0.5mm}$1$ & \hspace{0.5mm}$56.4^{***}$  \\ 
\hline \hline
\multicolumn{3}{l}{\footnotesize ***$p<0.001$, **$p<0.01$, *$p<0.05$}
\label{tab:nested-models}
\end{tabular}
\end{table}

\begin{table}[t]
\renewcommand{\arraystretch}{1.2}
\renewcommand{\tabcolsep}{4pt}
\caption{The threshold value ($\theta$)for each centrality measure as determined by maximization of the mutual information, its significance level measured by the chi-square test, and the probability of successfully inferring the persistence based on centrality ($\mathbb{P}_S$). Significant $p$ values are marked with an asterisk.}
\centering
\begin{tabular}{p{3.0cm}p{1.5cm}p{1.5cm}p{1.3cm}} \hline \hline
Centrality & \hspace{3mm}$\theta$  &  \hspace{2mm}$\chi^2$ & \hspace{2mm}$\mathbb{P}_S$\\ \hline
\hspace{3mm}Indegree 	  & $0.012$ & $18.29^{***}$ & $72\%$\\
\hspace{3mm}Outdegree	  & $0.012$ & $13.36^{***}$ & $68\%$\\
\hspace{3mm}Closeness     & $0.053$ & $28.48^{***}$ & $75\%$\\
\hline \hline
\multicolumn{3}{l}{\footnotesize ***$p<0.001$, **$p<0.01$, *$p<0.05$}
\label{tab:mutual-inf}
\end{tabular}
\end{table}

\subsection{Persistence as an indicator of centrality}\label{subsec:persistance_as_predictor}

To establish whether a student's position within the network at the beginning of the Spring semester is correlated with participation in MI in a preceding Fall semester, we used the network data from the first collection of the Spring 2015 (one section) and Spring 2016 (two sections) semesters. All three collections took place at the same time during the semester (i.e., week 3).

Initially, we looked at taking any section of MI-M as an independent variable (a binary yes or no predictor), i.e., centrality.spring $\sim$ persistence. Using simple linear regression, we found slightly positive correlation with persistence only for closeness [$F(1,216)=8.95$, adjusted $p=0.012$, adjusted $R^2=0.04$]. Because of a collinearity between persistence and section or instructor variables, we could not simply control for their effect in our model. Thus, to provide additional scrutiny, we expanded the participation variable to account for the section that students took. That revealed the significance of closeness [positive correlation, $F(2,145)=88.6$, adjusted $p<0.001$, adjusted $R^2=0.54$] and betweenness [negative correlation, $F(2,145)=4.51$, adjusted $p=0.036$, adjusted $R^2=0.05$]. The number of students who switched sections between semesters was low ($N=11$), and since it bore no statistical power, we decided to not control for that factor. Expanding the participation variable based on the instructor teaching the course (i.e., instructor A versus B) gave significant positive correlations for two out of five centralities: outdegree [$F(1,146)=6.39$, adjusted~$p=0.036$, adjusted ~$R^2=0.04$] and closeness [$F(1,146)=120.1$, adjusted~$p<0.001$, adjusted~$R^2=0.45$]. Also, expanding along the year when MI-M was taken (i.e., Fall 2014 versus Fall 2015) led to significant correlations: slightly negative for betweenness [$F(1,146)=9.05$, adjusted~$p=0.012$, adjusted~$R^2=0.05$] and positive for closeness [$F(1,146)=81.5$, adjusted~$p<0.001$, adjusted~$R^2=0.35$]. For all these cases we found no statistically significant improvement in fits for multiple linear regression models with students' demographic and grade information included.

Finally, since centralities at the end of the Fall semester are positively correlated with persistence, it is natural to ask whether there is any relationship connecting the centralities themselves between semesters. Specifically, we wanted to determine whether centralities at the beginning of Spring semesters can be predicted based on corresponding centralities at the end of Fall semesters. To account for other factors as possible predictors, we again used a backward elimination regression approach and we ended up with simple, one-predictor models with centrality alone giving the best fit. Out of five centralities we considered, three turned out to be statistically significant predictors. Table~\ref{tab:centr-centr} summarizes the regression results.

\begin{table}[t]
\renewcommand{\arraystretch}{1.1}
\renewcommand{\tabcolsep}{6pt}
\caption{Results of the linear regression for centralities at the beginning of Spring semesters as predicted by corresponding centralities at the end of Fall semesters (centrality.spring $\sim$ centrality.fall). Significant adjusted $p$ values are marked with an asterisk.}
\centering
\begin{tabular}{p{2.6cm}p{1.3cm}p{1.7cm}p{1.3cm}}
\hline \hline
Centrality.Fall 	& B & $F(1,146)$ & $R^2_{adj}$\\ \hline
Indegree 	& $0.13$ & \hspace{3.5mm}$2.6$ & $0.01$ \\ 
Outdegree 	& $0.37^{***}$ & \hspace{3.5mm}$20.6$ & $0.12$ \\ 
Eigenvector	& $0.11$ & \hspace{3.5mm}$2.2$ & $0.01$ \\ 
Betweenness & $0.21^{**}$  & \hspace{3.5mm}$11.1$ & $0.06$ \\
Closeness 	& $0.36^{***}$ & \hspace{2mm}$138.9$  & $0.48$ \\ 
\hline \hline
\multicolumn{3}{l}{\footnotesize ***$p<0.001$, **$p<0.01$, *$p<0.05$}
\label{tab:centr-centr}
\end{tabular}
\end{table}

\section{Discussion and conclusion}\label{sec:summary}

The use of SNA to study students' persistence was proposed by Tinto in the mid-1990s~\cite{Tinto97-CAC}. Thomas took this approach to explore the role of student social structure (e.g., integration) in persistence. He argued that the measures of centrality provide a unique empirical way to understand and quantify students' structural integration into their social groups. Following Tinto, he suggested that SNA sheds new light on understanding student integration through individual's social ties, i.e., ``a dimension that previous operationalizations of integration have missed''~\cite{Thomas00-TTB}.

As noted earlier, Tinto pointed out that both social and academic involvement influence persistence: ``The manner in which social and academic involvements (integration) shape learning and persistence will vary over the course of the college career and do so in differing ways for different students inside and outside the classroom''~\cite{Tinto97-CAC}. Our intro-sequence study looking at students at the beginning of their college career is put forward to improve the understanding of the role of social and academic interactions in persistence. To complement our findings, we are currently employing the SNA methodology to look at students' experiences at the university in the middle and at end of their time in college.

The MI physics course is interaction driven, at both the small and large group levels. The course structure---group assignments, common exam and lab reports, ``board meetings''---provides many active ways for students to be involved and make connections with each other. It is thus an important case to study the effect of involvement and/or integration on persistence.

In our analysis, we examined data from four semesters (six sections, three in Fall and three in Spring) taught by two instructors. To quantify various interpersonal interactions between students, we used centralities---measures of position within the social network. The first thing worth noticing is that the SNA measures are robust when compared between different groups. The network proper- ties, as well as correlations between centrality measures and persistence, remained fairly stable between years and sections. This stability confirms significance of our results.

Our preliminary work showed a positive correlation for persistence with directed degrees and closeness, and negative for betweenness~\cite{Zwolak15-SIP}. Building on this, we wanted to further investigate what information can be gained from centralities. We also wanted to go one step further and determine whether centrality at the beginning of the second semester can be predicted by measures from the end of the preceding semester.

Answering the first question, we verified that students with higher centralities at the end of the first semester of an MI course are in fact more likely to enroll in a second semester of MI physics. Both students who reported a large number of interactions and those who were often the subject of others' interactions were more likely to register for the second MI course in the introductory physics sequence. Node-level measures, i.e., directed degrees, and one of the whole network-level centralities, i.e., closeness, turned out to be positively correlated with persistence. However, betweenness had no statistically significant effect. This held true regardless of whether we included the instructional staff in our analysis or not. We also tested the impact of other factors, such as student demographics and final grades, and found grade to be the only one that made a statistically significant difference. However, further analysis showed that only for indegree was grade alone a better predictor. For outdegree and closeness, models with two dependent variables (i.e., grade and centrality) gave the best fit.

To explain these findings one needs to understand what centrality measures mean in the classroom context. Degree-based measures are concerned with communication activity. They capture students' direct interactions with one another. Indegree helps to identify individuals sought by others because, e.g., they are knowledgeable, supportive, or helpful. Outdegree, on the other hand, reveals students who reach out to others. They might do so because they need help or because they want to offer help and support. Some students like discussing their ideas to reaffirm their knowledge, others learn best from peer-to-peer interactions. The reasons for interacting with others are plenty. Understanding how and why students build communities is essential to improving their experiences in and beyond the classroom.

Concern with either independence or efficiency leads to the choice of a measure that captures embeddedness within the entire network, such as closeness. Students with high closeness scores have easy access to information from many sources. They are also---by sheer nature of this measure---connected to many students and might experience better social and emotional support. This, in turn, can help them appreciate all the benefits of having strong network connections within a classroom. Closeness is related to degrees of separation. Students who had low degree of separation from everyone else in the classroom were more likely to persist in the MI sequence of introductory physics courses.

Betweenness, on the other hand, depends mainly on the position within the network. In practice, in order to have high betweenness it suffices to connect groups that would otherwise be separate. Students with high betweenness score are not necessarily connected to many other students, but their connections are formed in a particular way. Thus, this measure will only be significant when the throughput or flow of information is relevant, but not necessarily when network connections are a source of support (which, e.g., implicitly or explicitly encourage students to persist).

We found that students who had a low degree of separation from others (i.e., a high closeness) and those that reached out to more peers were more likely to persist. This is true regardless of their grades in the course, which provides evidence that persistence through a major, in this particular context, does not depend solely on doing well academically, but also on doing well socially. Students' indegree did not predict their persistence when we took their grades into account. In other words, whether or not a student was perceived as a meaningful academic resource by their peers had no effect on their persistence, so long as they had high grades in the course.

Knowing the correlation between centralities and persistence leads to a natural question about the likelihood that students with indices within a certain range will actually continue their education. To tackle this problem we relied on the concept of entropy of a random variable. The mutual information gave us a threshold that optimizes predictability for each centrality. Then we derived the probability of correctly inferring the persistence based on centrality. Identifying students who are less likely to persist is particularly important when there is still time to take actions to help them through classroom interventions designed to further promote integration. Since in order to know students' grades one has to wait until after the end of the semester and the SNA data collection took place a couple of weeks before final exams, in our estimation we decided not to include the grade, even when it improved the model.

In our proposed model, we hypothesized that demographic information and the classroom context combined with centrality measures would predict persistence. However, with the exception of final grade, we found no impact of the personal factors in the logistic regression models. The reason for the unusual absence of this effect might be the atypical gender and ethnicity distribution for a science class, with females accounting for 46\% of the population and traditionally underrepresented ethnicities accounting for 81\%. As for the classroom context, all sections of our study followed the Modeling Instruction curriculum. Also, we assumed the faculty support and the teacher-to-student ratio factors to be comparable between sections since the review session and exams were coordinated and all students had access to teaching assistants and learning assistants. While our analysis would benefit from a comparison with a traditionally taught course, a previous study found no network development in a non-interactive classroom~\cite{Brewe10-PLC}. Thus, with the cost of additional data collection outweighing the potential gains, we decided to put off surveying non-MI students for the time being.

We were also interested in determining whether students' participation in Fall MI and their position within the classroom network by the end of the semester can say anything about their position in the network at the beginning of the following semester. Are students who already experienced the MI collaborative environment more likely to make connections? Do they tend to connect with people they already know from a previous semester? Based on our analysis, closeness turned out to be the most robust measure. We found positive correlations for this centrality regardless of whether we controlled for section, year, or instructor or if we treated as a variable simply taking the MI-M course (a binary ``yes or no'' predictor). In each case closeness was significantly correlated with the predicting factor. Moreover, each measure was positively correlated with its counterpart from the last collection of the Fall semester. It is also worth noting that ties involving students who took MI-M accounted for about 90\% of all interactions (78\% of which were initiated by returning students), while connections only between people new to the MI course constituted less than 10\%.

In conclusion, this study suggests that student social integration influences persistence and predicts their social integration in following courses (here, in MI-EM which follows MI-M). Without interventions to help students better connect to the social fabric of the classroom, we can expect patterns of success to go unchanged. Nevertheless, it is important to stress that there is no guarantee that increasing students' centrality by encouraging their interactions will lead to better persistence. Further study is needed to test for the effect of increasing embededness within the classroom network through, e.g., structured and purposeful mixing of students. Our findings, however, suggest that it will help. We found that different types of social integration are quantified by different measures of centrality and correlate with persistence differently---some greatly, some when subject to specific conditions (instructor, year), and some not at all. Network analysis is a tool that allows us to study the link between involvement and persistence in a quantitative, empirical, and meaningful way. It thus gives an access---in real time---to predictive data that may be useful to ``nip attrition in the bud'' and keep students enrolled and engaged in STEM fields.

\begin{acknowledgments} 
Supported by NSF PHY 1344247. We would like to thank two faculty members for facilitating data collection.
\end{acknowledgments}

\end{document}